\documentclass{kapproc} 
\usepackage{epsfig}
\kluwerbib

\normallatexbib 
\newcommand{\be}{\begin{equation}}
\newcommand{\ee}{\end{equation}}

\begin{document}

\articletitle{
Cosmological parameters from the clustering of AGN}

\author{Spyros Basilakos}
\affil{Astrophysics Group, Imperial College London, Blackett Laboratory, \\
Prince Consort Road, London SW7 2BW, UK}
\email{s.basilakos@ic.ac.uk}

\begin{abstract}

We attempt to put constraints on different cosmological and biasing
models by combining the recent clustering results of X-ray sources in
the local ($z\le 0.1$) and distant universe ($z\sim 1$).
To this end we compare the measured angular 
correlation function for bright \cite{AA} and faint 
\cite{VIK} {\em ROSAT} X-ray sources respectively
with those expected in three spatially flat cosmological models. Taking
into account the different functional forms of the 
bias evolution, we find that there are two cosmological models which match well the data. 
In particular, low-$\Omega_{\circ}$ 
cosmological models ($\Omega_{\Lambda}=0.7$) that contain either (i) high 
$\sigma_{8}^{\rm mass}=1.13$ or (ii) low 
$\sigma_{8}^{\rm mass}=0.9$ respectively with different bias behaviour, best reproduce the 
AGN clustering results. While $\tau$CDM models with different bias behaviour 
are ruled out at a high significance level.

\end{abstract}


\section*{Introduction}
The study of the distribution of matter on
large scales, based on different extragalactic objects,
provides important constraints on models of cosmic structure formation.
In particular Active Galactic Nuclei (AGN) can be detected up to very high redshifts
and therefore provide information on how the X-ray selected sources trace the 
underlying mass distribution as well as the evolution of large scale 
structure.

However, a serious problem here is how the luminous matter traces the 
underlying mass distribution. Many authors have claimed that the large scale 
clustering pattern of different mass tracers (galaxies or clusters) 
is characterized by a bias 
picture \cite{KA}. In this framework, biasing is assumed to be statistical in nature;
galaxies and clusters are identified as
high peaks of an underlying, initially Gaussian, random density field. 
Biasing of galaxies with respect to the dark matter distribution was also 
found to be an essential ingredient of CDM models of galaxy formation in 
order to reproduce the observed galaxy 
distribution. 
Furthermore, different studies have shown that the bias factor, $b(z)$,
is a monotonically increasing function of redshift 
\cite{TEG}. 
For example, Steidel et al. \cite{STEI} confirmed that the
Lyman-break galaxies are 
very strongly biased tracers of mass and they found that 
$b(z=3.4) \simeq 6, 4, 2$, for SCDM, $\Lambda$CDM $(\Omega_{\circ}=0.3)$ and
OCDM $(\Omega_{\circ}=0.2)$, respectively. 

Studies based on the traditional indicators of clustering, like the 
two point correlation function, have been utilized in order to 
describe the AGN clustering properties. Our knowledge
regarding the AGN clustering comes mostly
from optical surveys for QSO's (\cite{SHA} and reference therein).  
It has been established, that QSO's have 
a clustering length of $r_{\circ} \simeq 5.4\pm 1.1h^{-1}$Mpc. 

Vikhlinin \& Forman \cite{VIK} studied the 
angular clustering properties using a set of deep {\em ROSAT} observations.
Carrera et al. \cite{CA} combined two soft X-ray surveys (235 AGN), found a spatial correlation 
length $1.5h^{-1}$Mpc $\le r_{\circ} \le 5.5h^{-1}$Mpc, 
depending on the adopted model of clustering evolution. 
Recently, Akylas et al. \cite{AA} using  
2096 sources detected in the {\em ROSAT} All 
Sky Survey Bright Source Catalogue (RASSBSC), 
derived the AGN angular correlation function in the nearby Universe 
and utilizing Limber's equation obtained 
$r_{\circ}=6.7\pm 1.0h^{-1}$Mpc, assuming comoving clustering evolution.  

Here, we present the standard theoretical approach  
to estimate the angular correlation 
function $w(\theta)$, using different models for the bias evolution in 
different spatially flat cosmological models.
Comparing the latter with observational results, 
we attempt to put constraints on the different cosmological and bias
models

\section{The Integral Equation of Clustering}
For the purpose of this study we utilize the relation between the angular 
$w(\theta)$ and spatial $\xi(r,z)$ two point 
correlation functions (\cite{MAG} and references theirin).
As it is well known, this connection can be done using the Limber 
equation \cite{PEE}. For example, in the
case of a spatially flat Universe ($\Omega_{\circ}+\Omega_{\Lambda}=1$), the Limber 
equation can be written as

\begin{equation}\label{eq:angu}
w(\theta)=2\frac{H_{\circ}}{c} \frac{\int_{0}^{\infty} N^{2}(z)E(z){\rm d}z 
\int_{0}^{\infty} \xi(r,z) {\rm d}u}{[\int_{0}^{\infty} N(z) {\rm d}z]^{2}} \;\;.
\end{equation} 
where $x$ is the comoving coordinate related to the redshift through 
\be
x=\frac{c}{H_{\circ}} \int_{0}^{z} \frac{{\rm d}y}{E(y)} \;\; ,
\ee
with 
\be 
E(z)=[\Omega_{\circ}(1+z)^{3}+\Omega_{\Lambda}]^{1/2} 
\ee
\cite{PEE}. 
The mean surface density, ${\cal N}$, on a survey of solid angle $\Omega_{s}$ is:

\be
{\cal N}=\int_{0}^{\infty} x^{2} \phi(x) {\rm d}x=\frac{1}{\Omega_{s}} 
\int_{0}^{\infty} N(z) {\rm d}z \;\; ,
\ee
where $N(z)$ is the number of objects in the given survey within the shell $(z,z+{\rm d}z)$
and $\phi(x)$ is the selection function (the probability 
that a source at a distance $x$ is detected in the survey):
\be 
\phi(x)=\int_{L_{\rm min}}^{\infty} \Phi(L_{x},z) {\rm d}L \;\;.
\ee
In this work we used a luminosity function $\Phi(L_{x},z)$ 
of the form assumed by Boyle et al. \cite{BO}. 

The physical separation between two sources, 
separated by an angle $\theta$ considering 
a small angle approximation is given by:

\be
r \simeq \frac{1}{(1+z)} \left( u^{2}+x^{2}\theta^{2} \right)^{2} \;\; .
\ee
Extending this picture, 
we quantify the evolution of clustering with epoch 
presenting the spatial correlation function of the X-ray sources as  
\be
\label{eq:spat}
\xi(r,z)=\xi_{\rm mass}(r)R(z) \;\;, 
\ee
with $R(z)=D^{2}(z)b^{2}(z)$, where 
$D(z)$ is the linear growth rate of clustering (cf. Peebles 1993)
being given by   
\be\label{eq:24}
D(z)=\frac{5\Omega_{\circ} E(z)}{2}\int^{\infty}_{z} \frac{(1+y)}{E^{3}(y)} 
{\rm d}y 
\ee
and finally $b(z)$ is the evolution of bias. 

\section{Bias Evolution}
The concept of biasing between different classes of extragalactic objects 
and the background matter distribution was put forward by Kaiser \cite{KA}
in order to explain the higher amplitude of the 2-point correlation function 
of clusters of galaxies with respect to that of galaxies themselves.
Therefore, we shortly describe here
some of the bias evolution models in order to introduce them to our analysis. 

\subsection{No evolution of bias (B0)} 
This model considers constant bias at all epochs: 
\be
\label{eq:pres}
b(z)=b_{\circ} \simeq b_{\circ}^{\rm opt} \left( \frac{r_{\circ}}{r_{\circ}^{\rm opt}} \right)^{\gamma/2}
\ee
where $r_{\circ}^{\rm opt}=5.4h^{-1}$Mpc is the correlation length in comoving coordinates 
estimated by the APM correlation function \cite{MA} and 
$r_{\circ}$ is the corresponding length for 
the X-ray sources. Finally, $b_{\circ}^{\rm opt}=1/\sigma_{8}^{\rm mass}$ is the 
present bias of optical 
galaxies relative to the distribution of mass and $\sigma_{8}^{\rm mass}$ is the mass 
rms fluctuations in sphere of radius 8$h^{-1}$Mpc. Using the above ideas, 
if ones assumes that $r_{\circ} > r_{\circ}^{\rm opt}$, then it is quite 
obvious that the X-ray sources are indeed more biased with respect to optical
galaxies by the factor $(r_{\circ}/r_{\circ}^{\rm opt})^{\gamma/2}$. In case 
that $b(z)\equiv  1$, we have the so called 
non-bias model (from now on we consider $b(z)\equiv 1$ as a (B0) bias model).

\subsection{Test Particle or Galaxy Conserving Bias (B1)} 
This model, proposed by many authors 
(\cite{TEG} and references theirin), 
predicts the evolution of bias,
independent of the mass and the origin of halos, assuming only that the
test particles fluctuation field is related proportionally
to that of the underlying mass. 
Thus, the bias factor as a function of redshift can be written:
\be
b(z)=1+\frac{(b_{\circ}-1)}{D(z)} \;\; \;\; , 
\ee
where $b_{\circ}$ is the bias factor at the present time. 

\subsection{Merging Bias Model (B2)} 
Mo \& White \cite{MO} have developed a model for 
the evolution of the correlation bias, 
using the Press-Schechter formalism.
Utilizing a similar formalism, 
Matarrese et al. \cite{MAT} extended the Mo \& White
results to include the effects of different mass scales. In this 
case we have that

\be
b(z)=0.41+\frac{(b_{\circ} - 0.41)}{D(z)^{1.8}} \;\;.
\ee

\section{Cold Dark Matter (CDM) Cosmologies}

In this section, we present the cosmological models that we use in this work.
For the power spectrum of our CDM models, we consider $P(k) \approx k^{n}T^{2}(k)$ with
scale-invariant ($n=1$) primeval inflationary fluctuations. We utilize the transfer function 
parameterization as in Bardeen et al. \cite{BB}, with the corrections given 
approximately by Sugiyama's \cite{SU} formula:
$$T(k)=\frac{{\rm ln}(1+2.34q)}{2.34q}[1+3.89q+(16.1q)^{2}+$$
$$(5.46q)^{3}+(6.71q)^{4}]^{-1/4} \;\; .$$
with $q=k/h\Gamma$, where $k=2\pi/\lambda$ is the wavenumber in units of 
$h$ Mpc$^{-1}$ and $\Gamma$ is the CDM shape parameter in units 
of $(h^{-1} {\rm Mpc})^{-1}$. 
In this analysis, we have taken into account three different 
cold dark matter models (CDM) in order to isolate the effects of different parameters 
on the X-ray sources clustering predictions. 

The $\tau$CDM ($\Omega_{\circ}=1$, $h=0.5$, $\sigma_{8}^{\rm mass}=0.55$, $b_{\circ}=2.18$) 
and $\Lambda$CDM ($\Omega_{\Lambda}=0.7$, $h=0.65$, $\sigma_{8}^{\rm mass}=0.9$, $b_{\circ}=1.33$) 
\cite{MAR} are approximately
COBE normalized and the latter cosmological model
is consistent with the results from Type Ia supernova 
\cite{PER}. In the 
same framework, the $\tau$CDM and $\Lambda$CDM models 
have $\Gamma \sim 0.2$, in approximate agreement with 
the shape parameter estimated from galaxy 
surveys \cite{MA} and 
they have fluctuation amplitude in 8 $h^{-1}$Mpc scale, 
$\sigma_{8}^{\rm mass}$,  
consistent with the cluster abundance, 
$\sigma_{8}^{\rm mass}=0.55\Omega_{\circ}^{-0.6}$ \cite{EK}. 

Furthermore, in order to investigate cosmological models with high value of 
$\sigma_{8}^{\rm mass}$, we include a new model named $\Lambda$CDM2 
($\Omega_{\Lambda}=0.7$, $h=0.6$, $\sigma_{8}^{\rm mass}=1.13$, $b_{\circ}=1.06$) 
\cite{COL}. 
The $\sigma_{8}^{\rm mass}$ value for the latter cosmological model is in good 
agreement with both
cluster and the 4-years COBE data with a shape parameter $\Gamma=0.25$. Therefore, it 
is quite obvious that two of our  
models ($\tau$CDM and $\Lambda$CDM)  
have the same power spectrum and geometry but different values of $\Omega_{\circ}$ 
and $\sigma_{8}^{\rm mass}$, while the 
two spatially flat, low-density CDM models ($\Lambda$CDM and $\Lambda$CDM2)
have different $\sigma_{8}$ and $\Gamma$ respectively.  
It turns out that in $\Lambda$CDM and $\tau$CDM models the distribution of X-ray 
sources is ``biased'' relative to the distribution of mass; while the $\Lambda$CDM2 model is 
almost non ``biased'' while $b_{\circ}$ at the 
present time is described by equation (9). 

In order to understand better the effects of AGN clustering,
we present in figure 1 the quantity $R(z)=D^{2}(z)b^{2}(z)$ as a 
function of redshift for the three cosmological models, utilizing 
at the same time different bias evolution. It is quite
obvious that the behaviour of the function $R(z)$ characterizes the  
clustering evolution with epoch. Figure 1, for example, 
clearly shows that the bias at high redshifts has different values in different
cosmological models. In particular for the high 
$\sigma_{8}^{\rm mass}$ low-$\Omega_{\circ}$ flat model ($\Lambda$CDM2) 
the distribution of X-ray sources is only weakly
biased, as opposed to the strongly biased distribution in the 
$\tau$CDM cosmological model.       

Indeed the different functional forms of $b(z)$, provide clustering models 
where: (i) AGN clustering is a decreasing function with redshift for
(B0), (ii) AGN clustering is roughly constant for (B1). 
However, the $\Lambda$CDM2-B1 model gives lower $R(z)$
simply because the higher $\sigma_{8}^{\rm mass}$ normalization largely
removes the 
clustering difference between the two other flat cosmological models
with low $\sigma_{8}^{\rm mass}$ normalizations. In other words, the 
present bias value 
of the above model is almost $\sim 1$, which gives clustering 
behaviour similar to 
the $\Lambda$CDM2-B0. Finally, AGN clustering is a monotonically 
increasing function of redshift for (B2) and (B3) respectively
(\cite{BA} for details).  

\begin{figure}
\centerline{\epsfxsize=8truecm\epsfbox{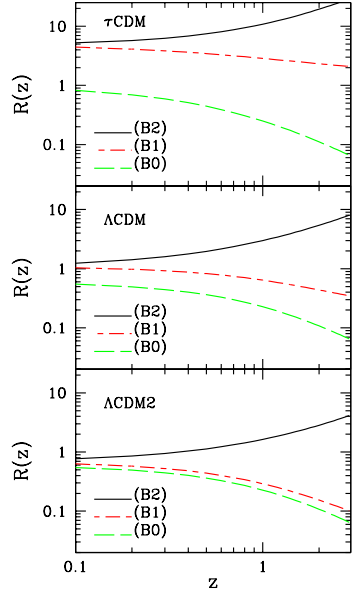}}
\caption{The function $R(z)$ as a function of $z$, 
for different bias evolution models.} 
\end{figure}

\section{Application to the Data}
In Figure 2 (left) we compare the angular correlation function of a 
sample of 2096 sources with a total sky coverage of 4.9sr detected in the 
{\em ROSAT} All-Sky Survey Bright Source Catalogue \cite{AA} with that predicted in various flat cosmological models. 
Considering a two point angular correlation function of the form
$w(\theta)=(\theta/\theta_{\circ})^{1-\gamma}$, 
the above authors
found $\theta_{\circ}=0.062^{\circ}$, $\gamma=1.8$ and spatial 
correlation length of $r_{\circ} \approx 6.5\pm 1.0h^{-1}$Mpc and 
$r_{\circ} \approx 6.7\pm 1.0h^{-1}$Mpc for stable and comoving clustering evolution 
respectively, similar to the optically selected AGN.
Due to the fact that the above estimates have been focused on the local X-ray Universe,  
in this work we use complementary observational results from
\cite{VIK}. 
They have analysed a set of deep {\em ROSAT} observations with a total sky
coverage of 40 deg$^{2}$, in order to investigate the clustering 
properties of faint X-ray sources. Therefore, they claimed that the two point angular 
correlation function is well described by a power low with
$\gamma=1.8$. Thus, in Figure 2 (right) 
we plot their results and the estimated angular correlation function for 
all nine models. 
\begin{figure}
\epsfxsize=7cm \epsffile{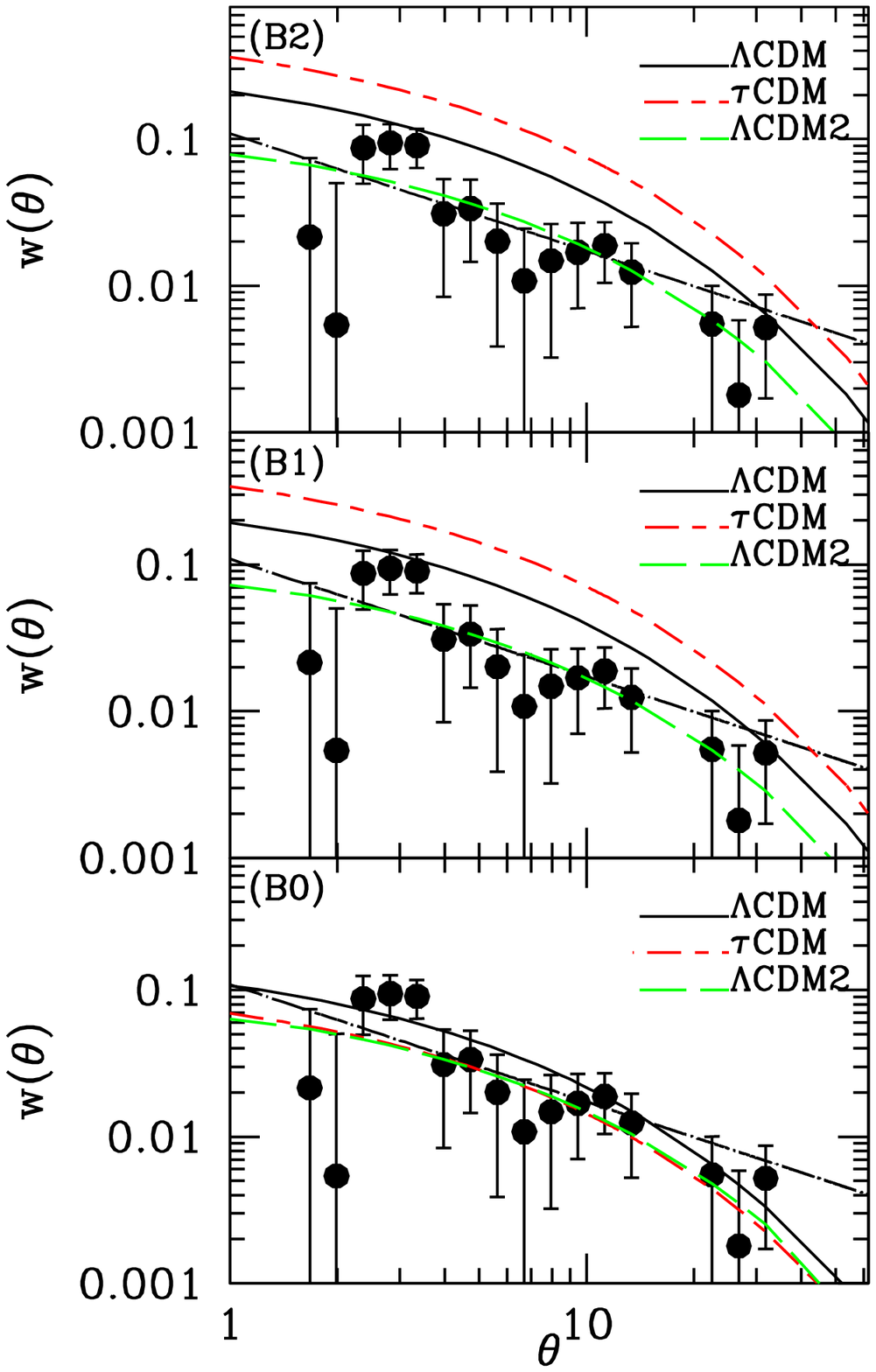} \hfill
\epsfxsize=7cm \epsffile{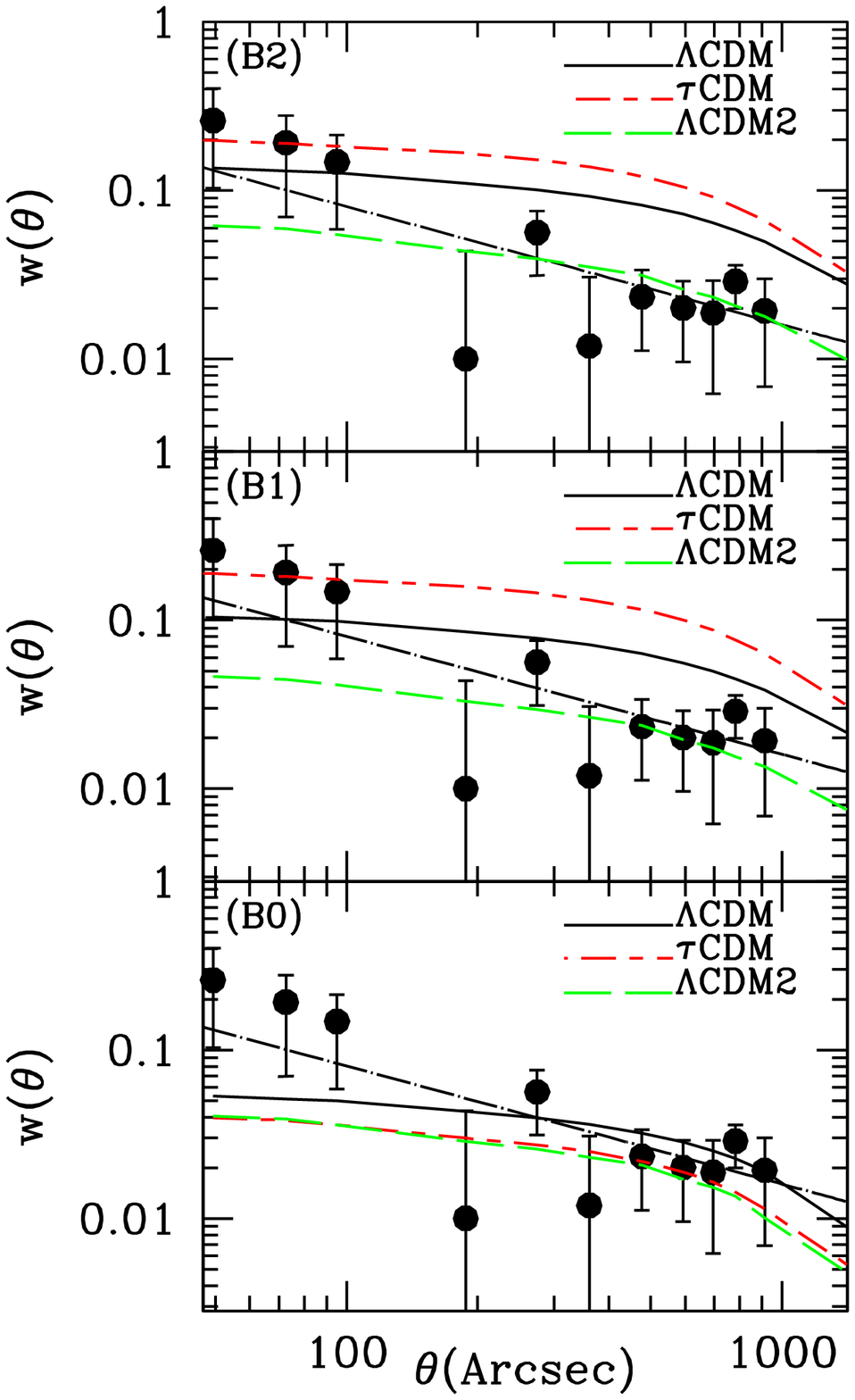}
\caption{Left: Comparison of the predicted angular correlation function 
for various cosmological models with that of the local AGN distribution, 
estimated in \cite{AA}. 
Errorbars are determined by assuming Poisson statistics. The continuous dot-dash line
represent the best fit to $w_{x}(\theta)$ derived by the above authors.
Right: Comparison of the predicted angular correlation function 
for various models with that estimated in \cite{VIK}.
The continuous dot-dash line
represent the best fit to $w_{x}(\theta)$ derived by the above authors.}
\end{figure}

\begin{table}
\caption[]{$\chi^2$ probabilities (${\cal P}_{>\chi^{2}}^{B, F}$)
of consistency between RASSBSC, faint {\em ROSAT} data and
models.} 
\tabcolsep 5pt
\begin{tabular}{cccc}
Comparison Pair &  ${\cal P}_{>\chi^{2}}^{B}$& Comparison Pair &  ${\cal P}_{>\chi^{2}}^{F}$\\  
RASSBSC - $\Lambda$CDM-B0 & 0.81   & faint {\em ROSAT} - $\Lambda$CDM-B0 & 0.037\\
RASSBSC - $\Lambda$CDM-B1 & 0.067 &faint {\em ROSAT} - $\Lambda$CDM-B1 & 3.67 $\times 10^{-4}$\\
RASSBSC - $\Lambda$CDM-B2 & 0.019 &faint {\em ROSAT} - $\Lambda$CDM-B2 & 1.72 $\times 10^{-6}$\\ \\
RASSBSC - $\Lambda$CDM2-B0 & 0.19& faint {\em ROSAT} - $\Lambda$CDM2-B0 & 3.44 $\times 10^{-4}$ \\
RASSBSC - $\Lambda$CDM2-B1 & 0.42 &faint {\em ROSAT} - $\Lambda$CDM2-B1 & 3.21 $\times 10^{-3}$ \\
RASSBSC - $\Lambda$CDM2-B2 & 0.57 &faint {\em ROSAT} - $\Lambda$CDM2-B2 & 0.036 \\ \\
RASSBSC - $\tau$CDM-B0 & 0.19  &faint {\em ROSAT} - $\tau$CDM-B0 & 4.64 $\times 10^{-3}$ \\
RASSBSC - $\tau$CDM-B1 & 2.88$\times 10^{-8}$ &faint {\em ROSAT} - $\tau$CDM-B1 & 1.07$\times 10^{-12}$ \\
RASSBSC - $\tau$CDM-B2 & 3.00$\times 10^{-9}$ &faint {\em ROSAT} - $\tau$CDM-B2 & 4.73 $\times 10^{-15}$ \\
\end{tabular}
\vspace{-0.5cm}
\end{table}
In order to quantify the
differences between models and data, we perform a standard 
$\chi^{2}$ test, for the bright (RASSBSC) and faint sources respectively,
and we present the ${\cal P}_{>\chi^{2}}^{N}$ (see table 2), where we have been use $N=B,F$ for bright 
(RASSBSC) and faint X-ray sources respectively.
Comparing the statistical results for 
both (a) high (faint sources) and (b) low (bright) redshift regimes,
we can point out that for the bright X-ray sources the
only models that are excluded by the data, at 
a relatively high significance level, are $\Lambda$CDM-B2, $\tau$CDM-B1 and $\tau$CDM-B2. 
Interestingly, for the faint objects the excluded models are 
$\Lambda$CDM-B1, $\Lambda$CDM-B2, $\Lambda$CDM2-B0, 
$\Lambda$CDM2-B1, $\tau$CDM-B0, $\tau$CDM-B1 and $\tau$CDM-B2. 
The above differences between the two kind of populations are to 
be expected simply because the cosmological 
evolution plays an important role on large scale structure clustering 
due to the fact that the high redshift objects are more biased tracers of 
the underlying matter distribution with respect to the low redshift
objects \cite{STEI}. 
Also, from the faint X-ray sources results we would like to point out that 
there is not a single $\tau$CDM model that fits the data.

If we make the reasonable assumption that there is no correlation between
the two X-ray populations, mostly due to the large distances
involved, we can consider the RASSBSC and faint {\em ROSAT} catalogues 
as being independent of each other. Under this assumption the previous
statistical tests can be considered as independent (as indeed verified by a KS
test).
Therefore, the joint (overall) probability can be given by 
the following expression:
$P={\cal P}_{>\chi^{2}}^{B} {\cal P}_{>\chi^{2}}^{F}
\left[1-{\rm ln} ({\cal P}_{>\chi^{2}}^{B}{\cal
P}_{>\chi^{2}}^{F})\right]\;.$ 
This overall statistical test proves that 
the $\Lambda$CDM-B0 ($P=0.14$) and $\Lambda$CDM2-B2 ($P=0.1$)   
models fit well the observational data at a
relatively high significance level. 

We should conclude that the behaviour of the observed angular 
correlation function of the X-ray sources is sensitive to the different cosmologies but 
at the same time there is a strong dependence on the bias models that 
we have considered in our analysis. By separating between low and 
high redshift regimes, we obtain results being consistent with the 
hierarchical clustering scenario, in which the 
AGN's are strongly biased at all cosmic epochs \cite{MAG}.

\section{Conclusions}
We have studied the clustering properties of 
the X-ray sources using the predicted angular correlation function 
for several cosmological models. We parametrize the predictions for $w(\theta)$ 
taking into account the behaviour of $b(z)$ for a non-bias model, 
a galaxy conserving bias model with $b(z) \propto (1+z)$ and  
for a galaxy merging bias model with $b(z) \propto (1+z)^{1.8}$. Utilising 
the measured angular correlation function, 
for faint and bright X-ray {\em ROSAT} sources,
estimated in \cite{VIK} 
and \cite{AA} respectively, we have compared them
with the corresponding ones predicted in three cosmological models,
namely the $\tau$CDM, $\Lambda$CDM and $\Lambda$CDM2 (with a high $\sigma_{8}^{\rm mass}$ value).  
We find that the models that best reproduce the observational results are: 
{\em (1)} $\Lambda$CDM2 model ($\Omega_{\Lambda}=1-\Omega_{\circ}=0.7$) with 
high $\sigma_{8}^{\rm mass}=1.13$ and bias evolution described by $b(z)
\propto (1+z)^{1.8}$ and
{\em (2)} $\Lambda$CDM model ($\Omega_{\Lambda}=1-\Omega_{\circ}=0.7$) with 
low $\sigma_{8}^{\rm mass}=0.9$ and bias evolution by $b(z) \equiv 1$.

\end{document}